\documentclass[12pt]{article}
\usepackage{epsfig}
\newcommand{\bea}{\begin{eqnarray}}
\newcommand{\eea}{\end{eqnarray}}
\newcommand{\be}{\begin{equation}}
\newcommand{\ee}{\end{equation}}
\newcommand{\MSbar}{\overline{\rm MS}}

\title{Linearly polarized gluon density in the rescaling model}

\author{N.A.~Abdulov$^{1}$, X. Chen$^{2,3}$, A.V.~Kotikov$^{1}$, A.V.~Lipatov$^{1,4}$}

\begin{document}

\maketitle

\begin{center}
  {\it $^{1}$Joint Institute for Nuclear Research, 141980, Dubna, Moscow region, Russia}\\
  {\it $^{2}$Institute of Modern Physics, Chinese Academy of Sciences, Lanzhou 730000, China}\\
{\it $^{3}$School of Nuclear Science and Technology, University of Chinese Academy of Sciences, Beijing 100049, China}\\
  {\it $^{4}$Skobeltsyn Institute of Nuclear Physics, Lomonosov Moscow State University, 119991, Moscow, Russia}

\end{center}

\vspace{0.5cm}

\begin{center}

{\bf Abstract }
       
\end{center}

\indent
The low-$x$ behavior of the linearly polarized gluon density $h_{g}(x,k_t^2, Q^2)$
in nuclei is studied in
the rescaling model at small transverse momentum $k_t$.


\vspace{1.0cm}


\vspace{1.0cm}

\section{Introduction} \indent

Determination of parton (quark and gluon) distribution functions (PDFs) in a proton  and nuclei
is a rather important task for modern high energy physics.
In particular, detailed knowledge on the gluon densities 
is necessary for experiments planned at the Large Hadron Collider (LHC) and future colliders, such as
Electron-Ion Collider (EIC), Future Circular hadron-electron Collider
(FCC-he), Electron-Ion Collider in China (EicC)
and Nuclotron-based Ion Collider fAcility (NICA)~\cite{Anderle:2021wcy,Chen:2020ijn,Abir:2023fpo,Arbuzov:2020cqg,SPD}.
For unpolarized cases, there are a distribution of unpolarized gluons, denoted as $f_g(x,Q^2)$, and a distribution of
linearly polarized gluons $h_g(x,k^2_t,Q^2)$, which corresponds to interference between $\pm 1$ gluon helicity
states\footnote{In the literature, other notations $f_1^g(x, Q^2)$ and $h_1^g(x,k_t^2,Q^2)$ or $h_1^{\perp g}(x,k_t^2,Q^2)$ 
are also widely used.}.
Compared to $f_g(x,Q^2)$, function
$h_g(x,k^2_t,Q^2)$ is currently poorly known\footnote{See also recent review~\cite{Boussarie:2023izj}).} in comparison with $f_g(x,Q^2)$
and depends on the gluon transverse momentum $k_t$ (so called Transverse Momentum Dependent, or TMD gluon density).
A theoretical upper bound for $h_g(x,k^2_t,Q^2)$ was obtained~\cite{Mulders:2000sh,Boer:2011kf}.

Previously, we have derived an analytical expressions for linearly polarized gluon density in a proton 
and investigated its behavior at low $x$~\cite{Abdulov:2023fas}.
Our analysis was based on the small-$x$ asymptotics for sea quark and gluon densities
calculated in the generalized {\it double asymptotic scale} (DAS) approach~\cite{BF1,Q2evo}
(see also~\cite{Rujula}) and was done with leading order (LO) accuracy.
In the present note we extend the consideration~\cite{Abdulov:2023fas} for nuclei.

As it is known,
the study of deep inelastic scattering (DIS) of leptons on nuclei reveals the appearance of a significant nuclear
effect, which excludes the naive idea of the nucleus as a system of quasi-free nucleons (see, for example,~\cite{Arneodo:1992wf,Rith:2014tma} for review).
This effect was first discovered by the European Muon Collaboration in the 
domain of valence quark dominance, so it was called the EMC effect~\cite{Aubert:1983xm}.
Currently, there are two main approaches to study the nuclear PDFs (nPDFs).
In the first, which is currently more common, nPDFs
are extracted from global fits
(see the recent review~\cite{Paakkinen:2022qxn} and references therein) to nuclear data using empirical
parametrization of their normalizations and numerical solution of Dokshitzer-Gribov-Lipatov-Altarelli-Parisi
(DGLAP) equations~\cite{DGLAP} to describe the corresponding QCD evolution (dependence on scale $Q^2$).
The second strategy is based on some nPDF models (see \cite {Kulagin:2016fzf} for more information). 
Here we follow the rescaling model~\cite{Jaffe:1983zw} based on the assumption~\cite{Close:1983tn} that the
effective size of gluon and quark confinement in the nucleus is greater than in the free nucleon.
Within the perturbative QCD, it was pointed out~\cite{Jaffe:1983zw,Close:1983tn}
that this confinement rescaling predicts that nPDFs and PDFs can be connected by simply scaling the argument $Q^2$
(see also a review ~\cite{Jaffe:2212}).
Thus, one can say that the rescaling model demonstrates the features inherent in both approaches: there are certain relationships between conventional and nuclear PDFs that arise as a result of
shifting the values of the kinematic variable $Q^2$ and, at the same time, both densities obey DGLAP equations.

Initially, the rescaling model was proposed for the domain of valence quarks dominance, $0.2\leq x\leq 0.8$,
where $x$ is the Bjorken variable.
Recently it was extended to a small $x$~\cite{Kotikov:2017mhk,Kotikov:2018ass}, where 
certain shadowing and antishadowing effects\footnote{The investigations of shadowing and antishadowing effects (see~\cite{Stodolsky:1966am,Nikolaev:1975vy} and~\cite{Nikolaev:1975vy}, 
respectively) have been started even before
experimental data~\cite{EuropeanMuon:1988tpw} were appeared (see~\cite{Nikolaev:1981dh} for an overview).}
were found for the sea quark and gluon densities.
Our main goal is to apply the rescaling model to linearly polarized gluon density
$h_g(x,k^2_t,Q^2)$ and show its nuclear modification for small $x$ values.

\section{Approach} \indent

At LO of perturbation theory the linearly polarized gluon density in a proton
$h_g(x,k^2_t,Q^2)$ at low $k_t$ values has the following form~\cite{Sun:2011iw}
(all PDFs are multiplied by $x$):
\bea
&&h_g(x,Q^2) \equiv \frac{xk_t^2}{2M^2}\, {h}^{\perp g}_1(x,k^2_t,Q^2)
=\frac{2a_s(Q^2)}{\pi M^2} \int_x^1 \frac{dx_1}{x_1}\left(1-\frac{x}{x_1}\right)
\Bigl[C_Af_g(x_1,Q^2) + \nonumber \\
  &&+ C_F f_q(x_1,Q^2)\Bigr]  + ...
\,, \label{h}
\eea
where $C_A=N_c$, $C_F=(N_c^2-1)/(2N_c)$ for the color $SU(N_c)$ group.
Here $M=0.938$ GeV is proton mass,
\be
a_s(Q^2)=\frac{\alpha_s(Q^2)}{4\pi}= \frac{1}{\beta_0\ln(Q^2/\Lambda^2_{\rm LO})}
\label{as}
\ee
is related with the conventional strong coupling constant $\alpha_s(Q^2)$ and $\beta_0=11-2f/3$ is the first coefficient
of QCD $\beta$-function in the $\MSbar$-scheme and $f$ is the number of active quarks.

Considering low $x$ asymptotics for gluon density $f_g(x,Q^2)$, we have~\cite{Q2evo}:
\begin{eqnarray}
f_g(x,Q^2) &=&
f_g^{+}(x,Q^2) + f_g^{-}(x,Q^2), \nonumber \\
f^{+}_g(x,Q^2) &=&
A_g^+ I_0(\sigma)
\; e^{-\overline d_{+} s} + O(\rho),~~ A_g^+= A_g + \frac{4}{9} \,A_q,~~
\nonumber \\
f^{-}_g(x,\mu^2) &=& A_g^{-} \,
e^{- d_{-} s} \, + \, O(x),~~ A_g^- =-\frac{4}{9} \,A_q\,,
	\label{8.02}
\end{eqnarray}
\noindent
where $A_g$ and $A_q$ are magnitudes of gluon and (sea) quark densities at some initial scale $Q_0^2$,
$I_0$ is modified Bessel function.
Here also
\be
s=\ln \left( \frac{a_s(Q^2_0)}{a_s(Q^2)} \right),~~
\sigma = 2\sqrt{\left|\hat{d}_+\right| s
  \ln \left( \frac{1}{x} \right)},~~
\rho=\frac{\sigma}{2\ln(1/x)},~~
\label{intro:1a}
\ee
and
\begin{equation}
\hat{d}_+ = - \frac{4C_A}{\beta_0} = - \frac{12}{\beta_0},~~~
\overline d_{+} = 1 + \frac{4f(1-C)}{3\beta_0} =
1 + \frac{20f}{27\beta_0},~~~
d_{-} = \frac{4Cf}{3\beta_0}= \frac{16f}{27\beta_0}
\label{intro:1b}
\end{equation}
are the singular and regular parts of the anomalous dimensions.

The linearly polarized gluon density in a proton
$h_g(x,Q^2)$, which is important mostly at low $x$ reqion,
has the following form~\cite{Abdulov:2023fas}
\begin{eqnarray}
h_g(x,Q^2) &=&
h_g^{+}(x,Q^2) + h_g^{-}(x,Q^2),~~ h^{-}_g(x,Q^2) = 0\,,  \nonumber \\
h^{+}_g(x,Q^2) &=&
\frac{2a_s(Q^2)}{\pi M^2} \left\{C_AA_g^+ \left(\frac{1}{\rho} \, I_1(\sigma)-I_0(\sigma)\right)
+C_FA_g^+ \left(I_0(\sigma)-\rho \, I_1(\sigma)\right) \right\}\; e^{-\overline d_{+} s}\,,
	\label{8.02h}
\end{eqnarray}
where $I_i$ is modified Bessel function.

\section{Rescaling model} \indent

In the rescaling model~\cite{Jaffe:1983zw},
the valence part of quark densities are changed in the case of
a $A$ nucleus at intermediate and large values of variable $x$ $( 0.2 \leq x \leq 0.8)$ as follows
\begin{equation}
  f_{V}^A(x,Q^2) =
  f_{V}(x,Q^2_{A,V}),
  \label{va.1}
\end{equation}
where the new scale $Q^2_{A,V}$ is related to $Q^2$ by~\cite{Kotikov:2017mhk}
\begin{equation}
s^A_V \equiv \ln \left(\frac{\ln\left(Q^2_{A,V}/\Lambda^2\right)}{\ln\left(Q^2_{0}/\Lambda^2\right)}\right)
= s +\ln\Bigl(1+\delta^A_V\Bigr) \approx s +\delta^A_V,~~~
\label{sA}
\end{equation}
i.e. the kernel modification of the main variable $s$ depends on the
$Q^2$-independent parameter $\delta^A_V$ having small values (see Tables 2 and 3 in~~\cite{Kotikov:2017mhk}).

\subsection{Rescaling model at low $x$} \indent

In~\cite{Kotikov:2017mhk}, the PDF asymptotics shown in (\ref{8.02}) were applied to the small $x$ region of the EMC
effect, using the simple fact that the sea quark and gluons densities increase with increasing $Q^2$.
Thus, in the case of nuclei, the PDF evolution scale is less than $Q^2$, and this can directly reproduce the
shadowing effect observed with global fits.
Since there are two components for each gluon density, see~(\ref{8.02}) and (\ref{8.02h}), we have two
free parameters $Q^2_{A,\pm}$, which can be determined from the analysis of experimental data for 
the EMC effect at low $x$ values.

Note that it is usually convenient to study the following ratio
\begin{equation}
R^{AD}_{t}(x,Q^2) = \frac{t^A_g(x,Q^2)}{t^D_g(x,Q^2)},
\label{AD}
\end{equation}
\noindent
for both conventional $f_g(x,Q^2)$ and linearly polarized $h_g(x,Q^2)$  gluon densities (here  $t = f, h$).
Taking advantage of the fact that the nuclear effect in the deuteron is very small\footnote{Study of nuclear 
effects for convetional PDFs in the deuteron can be found~\cite{AKP,Wang:2016mzo} (see also discussions~\cite{Kulagin:2016fzf}).}:
$t^D_g(x,Q^2) \approx t_g(x,Q^2)$,
we can assume that
\be
  t^{A}_g(x,Q^2)
~=~
t_g^{A,+}(x,Q^2) + t_g^{A,-}(x,Q^2),~~
t^{A,\pm}_g(x,Q^2) =
t^{\pm}_g(x,Q^2_{AD,\pm}).
\label{AD1}
\ee
\noindent
The expressions for $f^{\pm}_g(x,Q^2)$ and  $h^{\pm}_g(x,Q^2)$ are given in~(\ref{8.02}) and (\ref{8.02h}),
respectively, and the corresponding values of $s^{AD}_{\pm} $ turned out to be
\be
s^{AD}_{\pm} \equiv \ln \left(\frac{\ln\left(Q^2_{AD,\pm}/\Lambda^2\right)}{\ln\left(Q^2_{0}/\Lambda^2\right)}\right)
  = s +\ln\Bigl(1+\delta^{AD}_{\pm}\Bigr)\,,
\label{AD2}
\ee
where the results for $\delta^{AD}_{\pm}$ can be found~\cite{Kotikov:2017mhk}.

\section{Results} \indent

\begin{figure}
\centering
\vskip 0.5cm
\includegraphics[width=17.0cm]{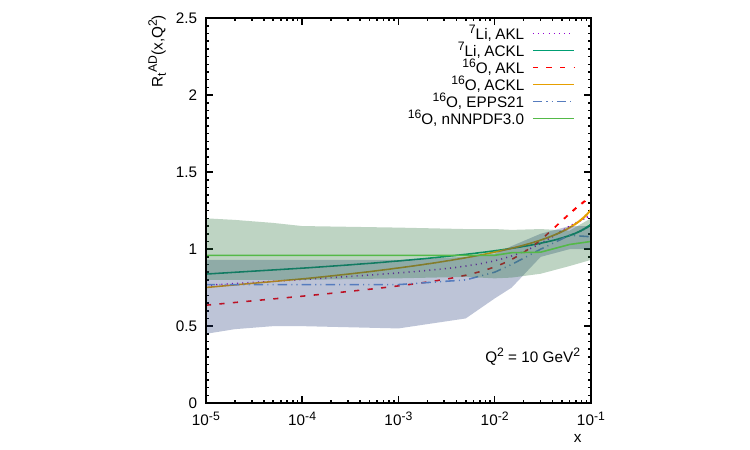}
\vskip -0.3cm
\caption{$x$ dependence of  $R^{AD}_{t}(x,Q^2)$
  at $Q^2$=10 GeV$^2$. 
The blue and yellow solid lines are for $R^{AD}_{h}(x,Q^2)$  obtained
in the present paper for ${}^7$Li and ${}^{16}$O, respectively.
The purple dotted and red dashed lines are for $R^{AD}_{g}(x,Q^2)$ 
obtained in~\cite{Abdulov:2022ypq}.
The blue dash-dotted and green lines and bands
are borrowed from~\cite{Paakkinen:2022qxn}.
}
\end{figure}

Our numerical results obtained for $R^{AD}_{g}(x,Q^2)$ and $R^{AD}_{h}(x,Q^2)$ are shown in Fig. 1. As it was discussed~\cite{Abdulov:2022ypq}, 
our predictions for $R^{AD}_{g}(x, Q^2)$ are very close to the ones~\cite{Kotikov:2017mhk} obtained 
at $x\leq 10^{-2}$, since the parameters $\delta_{\pm}^{AD}$ are taken from this article.
Moreover, in the low $x$ our calculations are also close to the recent results~\cite{Paakkinen:2022qxn} obtained by fitting
experimental data.
We see that predicted $R^{AD} _{h}(x, Q^2)$ are quite similar to the $R^{AD}_{g}(x, Q^2)$,
but they are less affected by nuclear effects. Indeed, in both cases: in the shadowing area ($x\leq 0.05$)
and in the antishadowing area ($0.05\leq x\leq 0.1$), the values of $R^{AD}_{h}(x, Q^2)$ less differ from 1
than the corresponding values of $R^{AD}_{g}(x, Q^2)$.
So that, the derived expressions could be useful for subsequent phenomenological applications.

In our forthcoming studies, we plan to extend our analysis for $x\geq 0.1$ using recently obtained PDF set~\cite{Abdulov:2022itv}. 
It is a combination of exact analytical solutions of DGLAP equations
for small and large values of $x$ and, therefore, could be applicable over the entire range of $x$.
We also plan to apply other modification models for
nPDFs~\cite{Arneodo:1992wf,Rith:2014tma,Paakkinen:2022qxn,Kulagin:2016fzf,Ma:2023tsi}.
Studies in a wide $x$ range could be supplemented with 
IMParton framework~\cite{Wang:2016sfq}. 
In addition, we will study nuclear modifications of transverse momentum dependent PDFs (or TMD gluon and quark densities)~\cite{Abdulov:2022itv,Kotikov:2019kci}, which
are  close relatives of  $h_g(x,k_t^2,Q^2)$. Now the these quantities become very popular (see ~\cite{Abdulov:2021ivr} and references and discussions therein).\\


Researches described in Section~2 were 
supported by the Russian Science Foundation under grant 22-22-00387.
Studies described and performed in Sections 3 and 4 were supported by the Russian Science Foundation under
grant 22-22-00119.

\end{document}